%% file: main.tex
\documentclass[11pt,a4paper]{article}

\usepackage[margin=2.5cm,top=1.5cm,bottom=1.5cm]{geometry}

\usepackage{amssymb}
\usepackage{amsmath}
\usepackage{breqn}
\usepackage{natbib}
\usepackage{hyperref}
\usepackage{scalefnt}
\usepackage{amsfonts}
\usepackage{graphicx,algorithmicx}
\usepackage{algorithm,algpseudocode}
\usepackage{setspace}

\usepackage{microtype}      
\usepackage{graphicx}
\usepackage{subcaption}
\usepackage{amsmath}
\usepackage{tabularx}
\usepackage{amssymb,amsthm}
\usepackage{enumitem}

\usepackage[font=small,labelfont=bf]{caption}

\usepackage{blindtext,mathtools}
\usepackage{multirow,float}
\usepackage{anyfontsize}
\usepackage{booktabs,geometry}
\usepackage{tabularx,array}
\usepackage{makeidx,url}
\usepackage{color,bbm,bm}

\usepackage{changes}
\usepackage{lipsum}
\definechangesauthor[name ={petros}, color = red]{PD}
\definechangesauthor[name ={mattia}, color = orange]{AA}
\definechangesauthor[name ={modification}, color = green]{MT}

\newcommand{\beqa}{\begin{eqnarray}}
\newcommand{\eeqa}{\end{eqnarray}}

\newcommand{\stdgausspi}
{\widetilde{\pi}_0} 

\newcommand{\gengausspi}
{\widetilde{\pi}} 

\newcommand{\solgauss}
{\hat{F}_{\widetilde{\pi}}}

\newcommand{\solgaussZero}
{\hat{F}_{\widetilde{\pi}_0}}

\newcommand{\stdgaussG}{G_0}
\newcommand{\gengaussG}{G}

\newtheorem{theorem}{Theorem}

\newtheorem{proposition}[theorem]{Proposition}
\newtheorem{remark}{Remark}
\newtheorem{assumption}{Assumption}

\if@twoside
\else
\fi
\def \dsp{\def \baselinestretch{0.9}\large \normalsize}
\dsp

\newcommand{\E}{\mathrm{E}}

\newcommand{\bX}{\mathbf{X}}
\newcommand{\R}{\mathbb{R}}
\usepackage{enumitem}

\providecommand{\keywords}[1]
{
  \small	
  \textbf{\textit{Keywords---}} #1
}

\title{\bf Variance Reduction for Metropolis-Hastings Samplers}

\author
{
    Angelos Alexopoulos
    \thanks{Corresponding author}\,
    \thanks{Medical Research Council Biostatistics Unit, University of Cambridge, Cambridge Institute of Public Health, Forvie Site, Robinson Way, Cambridge Biomedical Campus, Cambridge, CB2 0SR, UK. 
                Email: {\tt angelos@mrc-bsu.cam.ac.uk}.
        }
    \and    
         Petros Dellaportas
    \thanks{Department of Statistical Science, University College London, UK, The Alan Turing Institute, London, UK and Department of Statistics, AUEB, Greece.
                Email: {\tt p.dellaportas@ucl.ac.uk}.
        } 
  \and           
       Michalis K. Titsias
   \thanks{DeepMind, London, UK.
               Email: {\tt mtitsias@google.com}.
        }
   }

\date{
}

\begin{document}

\maketitle

\begin{abstract}
We introduce a general framework that constructs estimators with reduced variance for random walk Metropolis and Metropolis-adjusted Langevin algorithms. The resulting estimators require negligible computational cost and are derived in a post-process manner utilising all proposal values of the Metropolis algorithms.  Variance reduction is achieved by producing control variates through the approximate solution of the Poisson equation associated with the target density of the Markov chain. The proposed method is based on approximating the target density with a Gaussian and then utilising accurate solutions of the Poisson equation for the Gaussian case. This leads to an estimator that uses two key elements: (i) a control variate from the Poisson equation that contains an intractable expectation under the proposal distribution, (ii) a second control variate to reduce the variance of a Monte Carlo estimate of this latter intractable expectation. Simulated data examples are used to illustrate the impressive variance reduction achieved in the Gaussian target case and the corresponding effect when target Gaussianity assumption is violated. Real data examples on Bayesian logistic regression and stochastic volatility models verify that considerable variance reduction is achieved with negligible extra computational cost.

\end{abstract}


\input{Introduction}

\input{Methodology}

\input{Applications}

\input{Discussion}

\section{Supplemental material}

The R code for reproducing the experiments is available at \url{https://gitlab.com/aggelisalexopoulos/variance-reduction}.

\section*{ACKNOWLEDGEMENTS}

The second author acknowledges  financial support from The Alan Turing Institute under grant TEDSA2/100056.

\bibliographystyle{Chicago}
\bibliography{myrefs}

\newpage
\appendix

\input{appendix_supplement}

\end{document}

%% file: Introduction.tex
\section{Introduction}

Statistical methods for reducing the bias and the variance of estimators have played a prominent role in Monte Carlo based numerical algorithms.  Variance reduction via control variates has a long and well studied history introduced as early as the work of \cite{kahn1953methods}, whereas an early non-parametric estimate of bias, subsequently renamed jackknife and broadly used for bias reduction, was first presented by \cite{quenouille1956notes}.  However, the corresponding theoretical developments in the more complicated, but extremely popular and practically important, estimators based on  MCMC algorithms has been rather limited.  The major impediment is the fact that the MCMC estimators are based on ergodic averages of dependent samples produced by simulating a Markov chain. 

We provide a general methodology to construct control variates for any discrete time random walk Metropolis (RWM) and Metropolis-adjusted Langevin algorithm (MALA)  Markov chains that can achieve, in a post-processing manner and with a negligible additional computational cost, impressive variance reduction when compared to the standard MCMC ergodic averages.  Our proposed estimators are based on an approximate, but 
accurate, solution of the Poisson equation for a multivariate Gaussian target density of any dimension.  

Suppose that we have a sample of size $n$ from an ergodic Markov chain $\{X_n\}_{n \geq 0}$ with continuous state space $\bX \subseteq \R^d$, transition kernel $P$ and invariant measure $\pi$.  A standard estimator of the mean $\E_{\pi}[F] := \pi(F) = \int F d\pi$ of a real-valued function $F$ defined on $\bX$  under $\pi$ is the ergodic mean 
\begin{equation}
\nonumber
\mu_n(F) \coloneqq \frac{1}{n}\sum_{i=0}^{n-1}F(X_i).
\end{equation}
which satisfies, for any initial distribution of $X_0$, a central limit theorem of the form
\begin{equation}
\nonumber
\sqrt n \left[ \mu_n(F) - \pi(F) \right] = n^{-1/2} \sum_{i=0}^{n-1} 
\left[ F(X_i) - \pi(F) \right] 
\overset{D}{\to} N(0,\sigma^2_F), 
\end{equation}
with the asymptotic variance given by 
\begin{equation}
\nonumber
\sigma^2_F := \lim_{n \to \infty} n E_\pi \left[ \left( \mu_n(F) - \pi(F)\right)^2 \right].
\end{equation}
Interesting attempts on variance reduction methods for Markov chain samplers include the use of antithetic variables \citep{barone1990improving, green1992metropolis,craiu2005multiprocess}, Rao-Blackwellization \citep{gelfand1990sampling}, Riemann sums \citep{philippe2001riemann} or  autocorrelation reduction \citep{mira2000non, van2001art, yu2011center}.   

Control variates have played an outstanding role in the MCMC variance reduction quiver.  A strand of research is based on \cite{assaraf1999zero} who noticed that a Hamiltonian operator together with a  trial function are sufficient to construct an estimator with zero asymptotic variance.  They considered a Hamiltonian operator of Schr{\" o}dinger-type that led to a series of zero-variance estimators studied by \cite{valle2010new}, \cite{mira2013zero} and \cite{papamarkou2014zero}. The estimation of the optimal parameters of the trial function is conducted by ignoring the Markov chain sample dependency, an issue that was dealt with by \cite{belomestny2020variance} by utilizing spectral methods.  The main barrier for the wide applicability of zero-variance estimators is that their computational complexity increases with $d$, see  \cite{south2018regularised}. Another approach to construct control variates is a non-parametric version of the methods presented by \cite{mira2013zero} and \cite{papamarkou2014zero} which lead to the construction of control functionals \citep{oates2017control,barp2018riemannian,south2020semi}. 
Although their computational cost with respect to $d$ is low, their general applicability is prohibited due to the cubic computational cost with respect to $n$ \citep{south2018regularised,oates2019convergence} and the possibility to suffer from the curse of dimensionality that is often met in non-parametric methods \citep{wasserman2006all}.  Finally, \cite{hammer2008control} proposed constructing control variates by expanding the state space of the Metropolis-Hastings algorithm.

An approach which is closely related to our proposed methodology attempts to minimise the asymptotic variance $\sigma^2_F$.  This seems a hard problem since a closed form expression of $\sigma^2_F$ is not available and therefore a loss function to be minimised is not readily available; see, for example, \cite{flegal2010batch}.  However, there has been a recent research activity based on the following observation by  \cite{andradottir1993variance}. If a solution ${\hat F}$ to the Poisson equation for $F$ was available, that is if for every $x \in \bX$ 
\begin{equation}
\label{eq: PE}
 F(x) + P{\hat F}(x) - {\hat F}(x) = \pi(F)
\end{equation}
where 
\begin{equation}
PF(x) := E_x [ F(X_1)] := E_x [ F(X_1)| X_0 = x], \nonumber
\end{equation}
then one could construct a function equal to $F(x) + P{\hat F}(x) - {\hat F}(x)$
which is constant and equal to $\pi(F)$.  It is then immediate that a zero-variance and zero-bias estimator for $F$ is given by 
\begin{equation}
\label{eq:modif_erg_mean}
\nonumber
\mu_{n,{\hat F}}(F) \coloneqq \frac{1}{n}\sum_{i=0}^{n-1}\{F(X_i) + P{\hat F}(X_i) - {\hat F}(X_i)\}
\end{equation}
which can be viewed as an enrichment of the estimator $\mu_n(F)$ with the (optimal) control variate $P{\hat F} - {\hat F}$.    
Of course, solving (\ref{eq: PE}) is extremely hard for continuous state space Markov chains, even if we assume that 
$E_{\pi}[F]$ is known, because it involves solving a non-standard integral equation. 
Interestingly, a solution of this equation (also called the fundamental equation) produces zero-variance estimators suggested by \cite{assaraf1999zero} for a specific choice of Hamiltonian operator. 
One of the rare examples that (\ref{eq: PE}) has been solved exactly 
for discrete time Markov chains
is the random scan Gibbs sampler where the target density is a multivariate Gaussian density, see \cite{dellaportas2012control}, \cite{dellaportas2009notes}.  They advocated that this solution provides a good approximation to  (\ref{eq: PE}) for posterior densities often met in Bayesian statistics that are close to multivariate Gaussian densities.  Indeed, since direct solution of  (\ref{eq: PE}) is not available, approximating ${\hat F}$ has been also suggested 
by  \cite{andradottir1993variance}, \cite{atchade2005improving},  \cite{henderson1997variance}, \cite{meyn2008control}. 

\cite{tsourti2012variance} attempted to extend the work by \cite{dellaportas2012control} to RWM samplers.  The resulting algorithms produced estimators with lower variance but the computational cost required for the post-processing construction of these estimators counterbalance the variance reduction gains. We build on the work by \cite{tsourti2012variance} here but we differ in that (i) we build new, appropriately chosen to facilitate analytic computations, non-linear $d$-dimensional approximations to ${\hat F}(x)$ rather than linear combinations of $1$-dimensional functions and (ii) we produce efficient Monte Carlo approximations of the $d$-dimensional integral $P{\hat F}(x)$ so that no extra computation is required for its evaluation.
Finally, \cite{mijatovic2018poisson} approximate numerically the solution of (\ref{eq: PE}) for $1$-dimensional RWM samplers and \cite{mijatovic2019asymptotic} construct control variates for large $d$ by employing the solution of (\ref{eq: PE}) that is associated with the Langevin diffusion in which the Markov chain converges as $d \to \infty$ \citep{roberts1997geometric}; this requires very expensive Monte Carlo estimation methods so it is prohibited for realistic statistical applications.

We follow this route and add to this literature by extending the work of \cite{dellaportas2012control} and \cite{tsourti2012variance} to RWM and MALA algorithms by producing estimators for the posterior means of each co-ordinate of a $d$-dimensional target density with reduced asymptotic variance and negligible extra computational cost. 
Our Monte Carlo estimator to compute the expectation $\pi(F)$ makes use of three components: 
\begin{enumerate}[label=(\alph*)]
\item An approximation $G(x)$ to the solution of the Poisson equation associated with the target $\pi(x)$, transition kernel $P$ and  function $F(x)$. 
\item A construction of $G(x)$ based on firstly approximating $\pi(x)$ with a Gaussian 
density $\widetilde{\pi}(x) = \mathcal{N}(x|\mu, \Sigma)$, and then specifying $G(x)$ by an accurate approximation to the solution of the Poisson equation for the approximate target $\widetilde{\pi}(x)$.
\item An additional control variate,  referred to as \emph{static} control variate, that is based on the same Gaussian approximation $\widetilde{\pi}(x)$ and
allows to reduce the variance of a Monte Carlo estimator for the intractable expectation $P G(x)$.  
\end{enumerate}

In Section 2 we provide full details of the above steps. We start by discussing, in Section 2.1, how all the above ingredients are put together to eventually arrive at the general form of our proposed  estimator in equation \eqref{eq:estimator_final}.
In Section 3 we present extensive simulation studies that verify that our methodology performs very well with multi-dimensional Gaussian targets and it stops reducing the asymptotic variance when we deal with a multimodal $50$-dimensional target density  with distinct, remote modes. Moreover, we apply our methodology to real data examples consisting of a series of logistic regression examples with parameter vectors up to $25$ dimensions and two stochastic volatility examples with $53$ and $103$ parameters. In all cases we have produced estimators with considerable  variance reduction with negligible extra computational cost.

\subsection{Some notation} 

In the remainder of the paper  we use a simplified notation where  both $d$-dimensional random variables and their values are denoted by lower case letters, such as $x = (x^{(1)}, \ldots, x^{(d)})$ and where $x^{(j)}$ is the  $j$th dimension or coordinate, $j=1,\ldots,d$; the subscript $i$ refers to the $i$th sample drawn by using an MCMC algorithm, that is $x_i^{(j)}$ is the $i$th sample for the $j$th coordinate of $x$; the density of the $d$-variate Gaussian distribution with mean $m$ and covariance matrix $S$ is denoted by $\mathcal{N}(\cdot|m,S)$; for a function $f(x)$ we set $\nabla \eqqcolon ( \partial f/\partial x^{(1)},\ldots,\partial f/\partial x^{(d)})$; $I_d$ is the $d \times d$ identity matrix and the superscript $\top$ in a vector or matrix denotes its transpose; $||\cdot||$ denotes the Euclidean norm; all the vectors are understood as column vectors.

%% file: Methodology.tex
\section{Metrolopis-Hastings estimators with control variates from the Poisson equation}
\label{sec:GandPG}

\subsection{The general form of  estimators for arbitrary targets}
\label{sec:sec2_1}

Consider an arbitrary intractable 
target $\pi$ from which we have obtained a set of correlated samples by simulating a Markov chain with transition kernel $P$ obtained by a  Metropolis-Hastings kernel invariant to $\pi$.
To start with, assume a function $G(x)$. By following the observation of \cite{henderson1997variance} the function $PG(x) - G(x)$ has zero expectation with respect to $\pi$ because the kernel $P$ is invariant to $\pi$.  Therefore, given $n$ correlated samples  from the target, i.e.\ $x_i \sim \pi$ with $i=0,\ldots,n-1$, 
the following estimator is unbiased 
\begin{equation}
\label{eq:estimator_with_onlyG}
\mu_{n,G}(F) \coloneqq \frac{1}{n}\sum_{i=0}^{n-1}\{F(x_i) + \underbrace{P G(x_i) - G(x_i)}_{\text{Poisson control variate}}\}.
\end{equation}
For general Metropolis-Hastings 
algorithms the kernel $P$ is such that the expectation $PG(x)$ takes the form 
\begin{align}
\label{eq:PG2}
PG(x) 
& = \int P(x,d y) G(y) \nonumber \\
& = \int  \alpha(x,  y) q(y|x) G(y) dy  
+ \Big(1-\int \alpha(x,y) q(y|x)  dy \Big) G(x)  \nonumber \\
& = G(x) + \int\alpha(x,y)(G(y)-G(x))q(y|x)dy,
\end{align}
where 
\begin{align}
\label{eq:alpha}
\alpha(x,y) =  \min \big\{ 1,r(x,y) \big\},~~r(x,y) = \frac{\pi(y)q(x|y)}{\pi(x)q(y|x)}
\end{align}
and $q(y|x)$ is the proposal distribution. By substituting  \eqref{eq:PG2} back into  estimator
\eqref{eq:estimator_with_onlyG} we obtain 
\begin{equation}
\label{eq:estimator_with_onlyG2}
\mu_{n,G}(F) \coloneqq \frac{1}{n}\sum_{i=0}^{n-1}\left\{ F(x_i) +  \underbrace{ \int\alpha(x_i,y)(G(y)-G(x_i))q(y|x_i)dy }_{\text{Poisson control variate}} \right\}.
\end{equation}
To use this estimator 
we need to overcome two obstacles: (i) we need to specify the function $G(x)$ and (ii) we need to deal with the intractable integral associated with the control variate. 

Regarding (i) there is a theoretical \emph{best choice} which is to set $G(x)$ to the function $\hat{F}(x)$
that solves the Poisson equation,
\begin{equation}
\label{eq: PE_MH_exactpi}
\int\alpha(x,y)(\hat{F}(y)-\hat{F}(x))q(y|x)dy  = - F(x) + \pi(F), \ \text{for every} \ x \sim \pi,
\end{equation}
where we have substituted in the 
general form of the Poisson equation
from  \eqref{eq: PE} the Metropolis-Hastings kernel. 
For such optimal choice for $G$ the estimator in \eqref{eq:estimator_with_onlyG2}  has zero variance, i.e.\ it equals to the exact expectation $\pi(F)$.  
Nevertheless, getting $\hat {F}$ for general high-dimensional intractable targets is not feasible, and hence we need to compromise with an inferior choice for $G$ that can only approximate $\hat{F}$. To get such $G$, we make use of a Gaussian approximation to the intractable target, as indicated by the assumption below.
\begin{assumption} The target $\pi(x)$ is approximated by a multivariate Gaussian $\gengausspi(x) = \mathcal{N}(x|\mu, \Sigma)$ and the covariance matrix of the proposal  $q(y|x)$ is proportional to $\Sigma$.
\label{assumption1}
\end{assumption}
The main purpose of  the above assumption is to establish the ability to construct an efficient RWM or MALA sampler.  Indeed, it is well-known that efficient implementation of these Metropolis-Hastings samplers when $d>1$ requires that the  covariance matrix of  $q(y|x)$ should resemble as much as possible the shape of $\Sigma$.  In adaptive MCMC \citep{roberts2009examples}, such a shape matching is achieved during the adaptive phase where $\Sigma$ is estimated. If $\pi(x)$ is a smooth differentiable function, $\Sigma$ could be alternatively estimated by a gradient-based optimisation procedure and it is then customary to 
choose a proposal covariance matrix of the form $c^2 \Sigma$ for a tuned scalar $c$. 

We then solve the Poisson equation for the Gaussian approximation by finding the function $\solgauss (x)$ 
that satisfies,
\begin{equation*}
\label{eq: PE_MH_approxpi}
\int \widetilde{\alpha}(x,y)(\solgauss(y)-\solgauss(x))q(y|x)dy  = - F(x) + \gengausspi(F), \ \text{for every} \ x \sim \gengausspi.
\end{equation*}
It is useful to emphasize the difference between this new Poisson equation  and the original Poisson equation in \eqref{eq: PE_MH_exactpi}. This new equation involves 
the approximate Gaussian target $\gengausspi$ and the corresponding ``approximate'' Metropolis-Hastings transition kernel $\widetilde{P}$, which now has been modified so that the ratio
$\widetilde{\alpha}(x,y)$ is obtained 
by replacing the exact target $\pi$
with the approximate target 
$\gengausspi$ while the proposal $q(y|x)$ is also modified  if needed.\footnote{For the standard RWM algorithm $q(y|x)$ remains exactly the same, while for MALA it 
needs to be modified by replacing the gradient $\nabla \log \pi(x)$ with  $\nabla \log \gengausspi(x)$.} 
Clearly, this modification makes $\widetilde{P}$
invariant to $\gengausspi$. When $\gengausspi$ is a good approximation to $\pi$, we expect also $\solgauss$ to closely approximate the ideal function $\hat{F}$. Therefore, in our method we propose to set $G$ 
to $\solgauss$ (actually to an analytic approximation of $\solgauss$) and then use it in the 
estimator \eqref{eq:estimator_with_onlyG2}.

Having chosen $G(x)$, we now discuss the second challenge (ii), i.e. dealing with the intractable expectation $\int \alpha(x_i,y)(G(y)-G(x_i))q(y|x_i) d y$. 
Given that for any drawn sample 
$x_i$ of the Markov chain there is also a corresponding proposed sample $y_i$ that is generated from
the proposal, we can unbiasedly approximate the integral with a single-sample 
Monte Carlo estimate, 
\begin{equation*}
\label{eq:single-sample_Monte_Carlo}
\int \alpha(x_i,y)(G(y)-G(x_i))q(y|x_i) d y \approx \alpha(x_i,y_i)(G(y_i)-G(x_i)), \ y_i \sim q(y|x_i).
\end{equation*}
Although $\alpha(x_i,y_i)(G(y_i)-G(x_i))$ is a unbiased stochastic estimate of the Poisson-type control variate, it can have high variance that needs to be reduced. We introduce a second control variate based on some function $h(x_i,y_i)$, that correlates well with $\alpha(x_i,y_i)(G(y_i)-G(x_i))$,  
and it has analytic expectation
$\E_{q(y|x_i)} [h(x_i,y)]$. 
We refer to this control variate as \emph{static} since it involves 
a standard Monte Carlo problem  
with exact samples from the tractable proposal density $q(y|x)$. To construct $h(x_i,y)$ we rely again on the Gaussian approximation 
$\gengausspi(x) = \mathcal{N}(x|\mu, \Sigma)$ as we describe in 
Section
\ref{sec:static_control}. 

With $G(x)$ and $h(x,y)$ specified, 
we can finally write down the general form of the proposed estimator that can be efficiently computed only from the MCMC output samples $\{x_i\}_{i=0}^{n-1}$ and the corresponding proposed samples 
$\{y_i\}_{i=0}^{n-1}$: 
\begin{equation}
\label{eq:estimator_final}
\mu_{n,G}(F) \coloneqq \frac{1}{n}\sum_{i=0}^{n-1}\left\{ F(x_i) + \underbrace{ \alpha(x_i,y_i)(G(y_i)-G(x_i))  }_{\text{Stochastic Poisson control variate}} +  \underbrace{h(x_i,y_i) - \E_{q(y|x_i)} [h(x_i,y)]}_{\text{Static control variate}}
\right\}.
\end{equation}
In practice we use a  slightly modified version 
of this estimator by adding a set of adaptive regression coefficients $\theta_n$ to further reduce the variance 
following \cite{dellaportas2012control}; see Section \ref{sec:final_modified_combined_withDK}.

 \subsection{Approximation of the Poisson equation for Gaussian 
 targets 
 \label{sec:Gaussian}} 
 
 \subsubsection{Standard Gaussian case} 
 \label{ref:stdGaussian}
In this section we construct an analytical approximation to the exact solution of the Poisson equation for the standard Gaussian $d$-variate  target $\stdgausspi(x) = \mathcal{N}(x|0,I_d)$ and for the function $F(x)=x^{(j)}$ where $1 \leq j \leq d$. 
We use the function $F(x)=x^{(j)}$ in the remainder of the paper which corresponds to approximating the mean value $\E_{\pi}[x]$, while other choices 
of $F$ are left for future work. 
We denote the exact unknown solution by $\solgaussZero$ and the analytical approximation by $\stdgaussG$. Given this target and some choice for $\stdgaussG$ we  express the expectation in \eqref{eq:PG2} as
$$
P\stdgaussG(x) = \stdgaussG(x)(1-a(x)) + a_g(x),
$$ 
where
\begin{align}
\label{eq:PGGaussian1}
\alpha(x) & = \int \min \left\{1, \exp\left\{ -\tfrac{1}{2}(y^\top y - x^\top x)\right\} 
\frac{q(x|y)}{q(y|x)} \right\} 
q(y|x) dy, \\
\label{eq:PGGaussian2}
\alpha_g(x) & = \int \min \left\{1, \exp\left\{ - \tfrac{1}{2}(y^\top y- x^\top x)\right\} 
\frac{q(x|y)}{q(y|x)} \right\}
\stdgaussG(y) q(y|x) dy.
\end{align}
The calculation of $P\stdgaussG(x)$ reduces thus to the calculation of the integrals $a(x)$ and $a_g(x)$. In both integrals $x^\top x$ is just a constant since the integration is with respect to $y$. Moreover, the MCMC algorithm we consider is either RWM or MALA with proposal 
  \begin{equation}
 q(y|x)=N(y| rx,c^2I),
 \label{eq:qyx_std}
 \end{equation}
where $r=1$ corresponds to RWM and $r=1-c^2/2$ to MALA while $c>0$ is the step-size. Both $\alpha(x)$ and $\alpha_g(x)$ are expectations under the proposal distribution $q(y|x)$.

One  key observation is that for any dimension $d$, $y^\top y$ is just an univariate random variable 
with law induced by $q(y|x)$. Then, $y^\top y$ together with $\log \tfrac{q(x|y)}{q(y|x)}$ can induce an overall tractable univariate random variable so that the computation of $\alpha(x)$ in \eqref{eq:PGGaussian1} can be performed analytically. The computation of $\alpha_g(x)$ is more involved since it depends on the form of $\stdgaussG$. Therefore, we propose an approximate $\stdgaussG$ by first introducing a parametrised  family 
that leads to tractable and efficient closed form computation of $\alpha_g(x)$. In particular, we consider the following weighted sum of exponential functions 
\begin{align}
\label{eq:G0_lemma}
\sum_{k=1}^K w_k \exp\{\beta_k^\top x - \gamma_k (x-\delta_k)^\top (x-\delta_k)\},
\end{align}
where $w_k$ and $\gamma_k$ are scalars whereas $\beta_k$ and $\delta_k$ are $d$-dimensional vectors. It turns out that using the form in \eqref{eq:G0_lemma} for $G_0$ 
we can analytically compute the expectation $PG_0$ as stated in Proposition \ref{prop:PG}. The proof of this  proposition and the proofs of all remaining propositions and remarks presented throughout Section \ref{sec:GandPG} are given in the Appendix.

\begin{proposition}
\label{prop:PG}
Let $a(x)$ and $a_g(x)$ given by \eqref{eq:PGGaussian1} and \eqref{eq:PGGaussian2} respectively and $G_0$ in $a_g(x)$ to have the form in \eqref{eq:G0_lemma}. Then, 
$$
a(x) = \E_{f}\big[\min\big(1,\exp\big\{-\tfrac{c^2\tau^2(f-x^\top x/c^2)}{2}\big\}\big)\big],
$$
where $\tau^2=1$ in the case of RWM and $\tau^2=c^2/4$ in the case of MALA and $f$ follows the non-central chi-squared distribution with $d$ degrees of freedom and non-central parameter $x^\top x/c^2$, and
$$
a_g(x) = \sum_{k=1}^K A_k(x) \E_{f_{k,g}}\big[\min\{1, \exp\{-\tfrac{\tau^2s^2_k}{2}(f_{k,g}-x^\top x/s_k^2)\}\}\big],
$$
where $f_{k,g}$ follows the non-central chi-squared distribution with $d$ degrees of freedom and non-central parameter $m_k(x)^\top m_k(x)/c^2$ and
$
A_k(x) =(1+2c^2\gamma_k)^{-d/2} \exp\bigg\{-\frac{r^2x^\top x}{2c^2}-\gamma_k\delta_k^\top \delta_k + \frac{m_k(x)^\top m_k(x)}{2c^2(1+2\gamma_kc^2)} \bigg\},
$ 
$m_k(x) = \dfrac{rx + c^2(\beta_k+\gamma_k\delta_k)}{1+2c^2\gamma_k}$ and $s_k^2=c^2/(1+2c^2\gamma_k)$.
\end{proposition}
Proposition \ref{prop:PG} states that the calculation of $a_g(x)$ and $a(x)$ is based on the cdf of the non-central chi-squared distribution and allows, for $d$-variate standard normal targets, the exact computation of the modified estimator $\mu_{n,G}$ given by  \eqref{eq:estimator_with_onlyG}. 

Having a family of functions for which we can calculate analytically the expectation $PG_0$ we turn to the problem of specifying a particular member of this family to serve as an accurate approximation to the solution of the Poisson equation for the standard Gaussian distribution. We first 
provide the following proposition which states that $\solgaussZero$ satisfies certain symmetry properties.

\begin{proposition}
\label{prop:properties_of_stdgausssol}
Given $F(x) = x^{(j)}$, the exact solution $\solgaussZero(x)$ is: (i) (holds for $d \geq 1$) Odd function in the dimension $x^{(j)}$. (ii) (holds for $d \geq 2$) Even function over any remaining dimension $x^{(j')}, j' \neq j$. (iii) (holds for $d \geq 3$) Permutation invariant over the remaining dimensions.  
\end{proposition}
To construct an approximation model family that incorporates the symmetry properties of Proposition 
\ref{prop:properties_of_stdgausssol} we make the following assumptions for the parameters in \eqref{eq:G0_lemma}. We set $K=4$ and we assume that $w_k \in \R$ and $\gamma_k >0 $ for each $k=1,2,3,4$ whereas we set $w_1=-w_2=b_0$, $w_3=-w_4=c_0$, $\gamma_1=\gamma_2 =b_2$ and $\gamma_3=\gamma_4 =c_1$.  Moreover, for the $d$-dimensional vectors $\beta_k$ and $\delta_k$ we assume that $\beta_1 = -\beta_2$, $\beta_3 = \beta_4 =\delta_1 = \delta_2 =0$ and $\delta_3 = -\delta_4$; we set the vectors $\beta_1$ and $\delta_3$ to be filled everywhere with zeros except from their $j$th element which is equal to $b_1$ and $c_2$ respectively. We specify thus the function $\stdgaussG:\R^d \rightarrow \R$ as 
\begin{equation}
\label{eq:G2dim}
G_0(x)  = b_0 (e^{b_1 x^{(j)}}-e^{-b_1 x^{(j)}})\times e^{-b_2 ||x||^2} + c_0 (e^{- c_1 (x^{(j)}-c_2)^2} - e^{- c_1(x^{(j)}+c_2)^2}) \times e^{-c_1 \sum_{j' \neq j} (x^{(j')})^2}.
\end{equation}

\begin{figure}[H]
    \centering
    \includegraphics[scale=0.5]{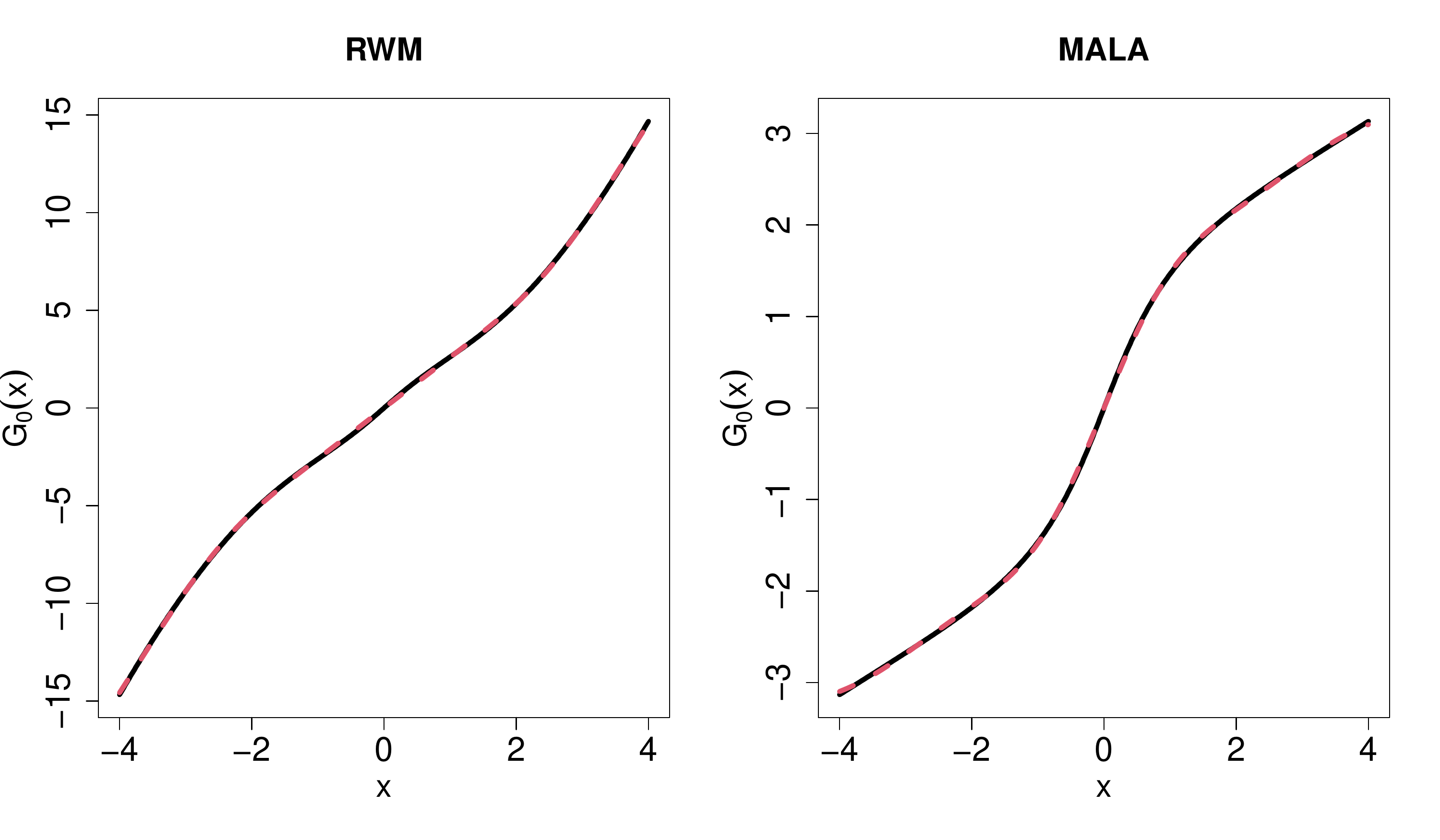}
    \caption{Numerical solution of the Poisson equation (black solid lines) and its approximation (red dashed lines) in the case of univariate standard Gaussian target simulated by using the random walk Metropolis (RWM) algorithm and the Metropolis-adjusted Langevin algorithm (MALA). 
    }
    \label{fig:G_1d_approx}
\end{figure}

To identify optimal parameters for the function $G_0$ in \eqref{eq:G2dim} such that $G_0 \approx \solgaussZero$ we first simulate a Markov chain with large sample size $n$ from the $d$-variate standard Gaussian distribution by employing the RWM algorithm and the MALA. Then, for each algorithm we minimize the loss function
\begin{equation}
\label{eq:loss}
    \mathcal{L} = (1/n)\sum_{i=1}^n(G_0(x_i)-PG_0(x_i)-x^{(1)}_i)^2,
\end{equation}
with respect to the parameters $b_0$, $b_1$, $b_2$, $c_0$, $c_1 $ and $c_2$ by employing the Broyden–Fletcher–Goldfarb–Shanno method. Figure \ref{fig:G_1d_approx} provides an illustration of the achieved approximation to $\solgaussZero$ in the univariate case where $d=1$ and
the model in \eqref{eq:G2dim} simplifies as 
\begin{equation*}
\label{eq:G1dim}
G_0(x)  = b_0 (e^{b_1 x}-e^{-b_1 x})\times e^{-b_2 x^2} + c_0 (e^{- c_1 (x-c_2)^2} - e^{- c_1(x+c_2)^2}).
\end{equation*}
For such case, we can visualize our optimised $G_0$ 
and compare it against the numerical solution from \cite{mijatovic2018poisson}. Figure \ref{fig:G_1d_approx} shows this comparison 
which provides clear evidence that for $d=1$ our approximation is very accurate.

 \subsubsection{General Gaussian case} 
 
 Given the general $d$-variate Gaussian target $\gengausspi(x) = \mathcal{N}(x|\mu,\Sigma)$ we
 denote by $\solgauss$ the exact solution of the Poisson equation and by $\gengaussG$ the approximation that we wish to construct.   
 To approximate $\solgauss$ 
 we apply a change of variables transformation from the 
 standard normal, as motivated by 
 the following proposition and remark.

\begin{proposition}
\label{prop:from_std_to_general_gauss}
Suppose the standard normal target $\stdgausspi(x) = \mathcal{N}(x|0,I_d)$, the function $F(x) = x^{(1)}$ and $\solgaussZero$ the associated solution of the Poisson equation for 
either RWM with proposal $q(y|x) = \mathcal{N}(y|x, c^2I)$ or MALA with proposal $q(y|x) = \mathcal{N}(y| (1 - c^2/2)x, c^2I)$. Then, the solution $\solgauss$ for the general Gaussian target $\gengausspi(x) = \mathcal{N}(x|\mu,\Sigma)$ and Metropolis-Hastings proposal
 \begin{equation}
 \label{eq:qyx}
 q(y|x)=
 \begin{cases} 
       \mathcal{N}(y| x,c^2\Sigma) & \text{if RWM} \\
      \mathcal{N}(y|  x +  (c^2/2)\Sigma\nabla \log \tilde{\pi}(x),c^2\Sigma) & \text{if MALA},  
   \end{cases}
 \end{equation}
is 
$
\solgauss(x) = L_{1 1} \solgaussZero(L^{-1}(x-\mu)),
$  
where $L$ is a lower triangular Cholesky matrix such that $\Sigma=L L^T$ and $L_{11}$ is its first diagonal element.
\end{proposition}
\begin{remark}
\label{rem:permutation}
To apply Proposition \ref{prop:from_std_to_general_gauss} for $F(x) = x^{(j)}$, $j\neq 1$, the vector $x$ needs to be permuted such that $x^{(j)}$ becomes its first element; the corresponding permutation has also to be applied to the mean $\mu$ and covariance matrix $\Sigma$.  \end{remark}

Proposition \ref{prop:from_std_to_general_gauss}
implies that we can obtain the exact solution of the Poisson equation 
for any $d$-variate Gaussian target by applying a change of variables transformation 
to the solution of the 
standard normal $d$-variate target. 
Therefore, based on  this theoretical result we propose to obtain an approximation $G$ of the Poisson equation in the general  Gaussian case by simply 
transforming the approximation $\stdgaussG$ in \eqref{eq:G2dim} from the standard normal case so as
\begin{equation}
\gengaussG(x) = \stdgaussG(L^{-1}(x-\mu)).
\label{eq:generalG}
\end{equation}
The constant $L_{1 1}$ is omitted since it can be absorbed by the regression coefficient $\theta$; see Section \ref{sec:final_modified_combined_withDK}.

 \subsection{Construction of the static control variate $h(x,y)$}
 \label{sec:static_control}

Suppose we have constructed 
a Gaussian approximation 
$\gengausspi(x) = N(x|\mu,\Sigma)$, where $\Sigma= L L^\top$,
to the intractable 
target $\pi(x)$ and also have obtained the function 
$\gengaussG$ from \eqref{eq:generalG} needed for the proposed, general, estimator in \eqref{eq:estimator_final}.
What remains is to specify the function 
$h(x,y)$, labelled as \textit{static control variate} in \eqref{eq:estimator_final}, which should correlate well 
with 
$
\alpha(x,y)(\gengaussG(y)- \gengaussG(x)).
$
The intractable term in this function is the Metropolis-Hastings probability $\alpha(x,y)$ in (\ref{eq:alpha})
where the Metropolis-Hastings ratio 
$r(x,y)$
contains the intractable target $\pi$. This suggests to choose $h(x,y)$ as 
\begin{equation}
    \label{eq:hxy}
    h(x,y)=\min\{1,\tilde{r}(x,y)\}\big[\gengaussG(y)- \gengaussG(x)\big],
\end{equation}
where $\widetilde{r}(x,y)$ is the acceptance ratio in a M-H algorithm that targets the Gaussian approximation $\gengausspi(x)$, that is
\begin{equation}
\label{eq:tilde_ratio}
    \widetilde{r}(x,y)=\min\bigg\{1,\frac{\gengausspi(y)\widetilde{q}(x|y)}{\gengausspi(x)\widetilde{q}(y|x)} \bigg\},
\end{equation}
and $\widetilde{q}(\cdot|\cdot)$ is the proposal distribution that we would use for the Gaussian target $\gengausspi(x)$ as defined by equation \eqref{eq:qyx}. Importantly, by assuming that $\widetilde{\pi}$ serves as an accurate approximation to $\pi$, the ratio $\widetilde{r}(x,y)$ approximates accurately the exact M-H ratio $r(x,y)$ and $\E_q[h(x,y)]$ can be calculated analytically. In particular, using \eqref{eq:generalG} we have that  
\begin{align*}
    \E_q[h(x,y)] & = \int h(x,y)q(y|x)d y\\
    &= \int \min\{1,\widetilde{r}(x,y) \}\big[\stdgaussG(L^{-1}(y-\mu))- \stdgaussG(L^{-1}(x-\mu)) \big] q(y|x) d y.
\end{align*}
This integral can be computed efficiently as follows. 
We reparametrize the integral according to the new variable $\tilde{y} = L^{-1} (y - \mu)$ and also use the shortcut $\tilde{x} = L^{-1}(x - \mu)$ where $x$ is an MCMC sample. After this reparametrization, the above expectation becomes under the distribution
 \begin{equation}
 \label{eq:qyx_reparam}
 q(\tilde{y}|\tilde{x})=
 \begin{cases} 
       \mathcal{N}(\tilde{y}| \tilde{x},c^2I) & \text{if RWM} \\
      \mathcal{N}(\tilde{y} | \tilde{x} + \frac{c^2}{2} L^\top \nabla \log \pi(x), c^2 I) & \text{if MALA},
   \end{cases}
 \end{equation}
where we condition on $\tilde{x}$ with a slightly abuse of notation since the term $\nabla \log \pi(x)$ is the exact pre-computed gradient for the sample $x$ of the intractable target. Thus, the calculation of $\E_q[h(x,y)]$ reduces to the evaluation of the following integral 
\begin{align}
\label{eq:Eq_static}
    \int \min\left\{1,\exp\{- \frac{1}{2}(\tilde{y}^\top \tilde{y} - \tilde{x}^\top \tilde{x})  \}\frac{\widetilde{q}(\tilde{x}|\tilde{y})}{\widetilde{q}(\tilde{y}|\tilde{x})} \right\} \big[\stdgaussG(\tilde{y})- \stdgaussG(\tilde{x}) \big]q(\tilde{y}|\tilde{x})  d \tilde{y}.
\end{align}
Note also that inside the Metropolis-Hastings ratio $\widetilde{q}(\tilde{y}|\tilde{x}) = \mathcal{N}(\tilde{y}|r\tilde{x},c^2I)$ with $r$ as in \eqref{eq:qyx_std}. In the case of RWM and by noting that the density $q(\tilde{y}|\tilde{x})$ in  \eqref{eq:qyx_reparam} coincides with the density $\widetilde{q}(\tilde{y}|\tilde{x})$ in \eqref{eq:qyx_std} we have that the calculation of the integral in \eqref{eq:Eq_static} reduces to the calculation of the integrals in \eqref{eq:PGGaussian1} and \eqref{eq:PGGaussian2} and, thus, can be conducted by utilizing Proposition \ref{prop:PG}. The calculation of the integral in \eqref{eq:Eq_static} for the MALA 
is slightly different as highlighted by the following remark.
\begin{remark}
\label{rem:MALA_static}
In the case of MALA the mean of the density $q(\tilde{y}|\tilde{x})$ in \eqref{eq:qyx_reparam} is different from the mean of $\widetilde{q}(\tilde{y}|\tilde{x})$ due to the presence of the term $\frac{c^2}{2} L^\top \nabla \log \pi(x)$ and the formulas in Proposition \eqref{prop:PG} are modified accordingly.
\end{remark}

Finally, we note that except from the tractability in the calculations which offered by the particular choice of $h(x,y)$, there is also the following intuition for its effectiveness. If the Gaussian approximation
is exact, then the overall control variate,  defined in equation \eqref{eq:estimator_final} as the sum of a stochastic and a static control variate,   
becomes the exact ``Poisson control variate'' that we would compute if the initial target was actually Gaussian. Thus, we expect that the function $h(x,y)$, as a static control variate in a non-Gaussian target, enables effective variance reduction under the assumption that the target is well-approximated by a Gaussian distribution.

 \subsection{The modified estimator with regression coefficients}
 \label{sec:final_modified_combined_withDK}
  
 As pointed out by \cite{dellaportas2012control} the fact that the proposed estimator $\mu_{n,G}(F)$ is based on an approximation $G$ of the true solution $\hat{F}_{\pi}$ of the Poisson equation implies that we need to modify $\mu_{n,G}(F)$ as
 \begin{align}
\label{eq:estimator_final_theta}
\mu_{n,G}(F_{\hat{\theta}_n}) &\coloneqq \frac{1}{n}\sum_{i=0}^{n-1}\left\{ F(x_i) + \hat{\theta}_n\big\{\underbrace{ \alpha(x_i,y_i)(G(y_i)-G(x_i))  }_{\text{Stochastic Poisson control variate}} +  \underbrace{h(x_i,y_i) - \E_{q(y|x_i)} [h(x_i,y)]}_{\text{Static control variate}}\big\}
\right\} 
\end{align}
where $\hat{\theta}_n$ estimates the optimal coefficient $\theta$ that further minimizes the variance of the overall estimator.  
\cite{dellaportas2012control} show that for reversible MCMC samplers, the optimal estimator $\hat{\theta}_n$ of the true coefficient $\theta$ can be constructed solely from the MCMC output.  By re-writing  the estimator in   \eqref{eq:estimator_final_theta} as
\begin{align}
\label{eq:estimator_final_theta2}
\nonumber
\mu_{n,G}(F_{\hat{\theta}_n}) & 
\coloneqq \frac{1}{n}\sum_{i=0}^{n-1}\{F(x_i) - \hat{\theta}_n \{G(x_i) -  \widehat{PG}(x_i) \} \},
\end{align}
where the term   
\begin{equation}
\label{eq:PGhat}
\widehat{PG}(x_i) =  G(x_i)+a(x_i,y_i)(G(y_i)-G(x_i))+ h(x_i,y_i)- E_{q(y|x_i)}[h(x_i,y)],
\end{equation}
approximates $PG(x_i)$, we can estimate $\hat{\theta}_n$ as
\begin{equation}
    \label{eq:theta_empirical}
    \hat{\theta}_n = \frac{\mu_n(F(G+\widehat{PG}))-
    \mu_n(F)\mu_n(G+\widehat{PG})}{\tfrac{1}{n}\sum_{i=1}^{n-1}\big(G(x_i)-\widehat{PG}(x_{i-1}) \big)^2 }.
\end{equation}
The resulting estimator $\mu_{n,G}(F_{\hat{\theta}_n})$ in \eqref{eq:estimator_final_theta} is evaluated by using solely the output of the MCMC algorithm and under some regularity conditions converges to $\pi(F)$ a.s. as $n \to \infty$, see \cite{tsourti2012variance}.

 \subsection{Algorithmic summary}
 \label{sec:final_algorithm}
 
 In summary, the proposed variance reduction approach can be applied 
 a posteriori to the MCMC output 
 samples $\{x_i\}_{i=0}^{n-1}$ obtained from either RWM or MALA with proposal density given by \eqref{eq:qyx}. The extra
 computations needed involve the evaluation  of $\widehat{PG}(x_i)$ given by  \eqref{eq:PGhat}. This is efficient since it relies on quantities that are readily available such as the values $G(x_i)$ and $G(y_i)$, where $y_i$ is the value generated from the proposal $q(y|x_i)$ during the main MCMC algorithm, as well as on the acceptance probability $a(x_i,y_i)$ which has been also computed and stored at each MCMC iteration. The evaluation of $\widehat{PG}(x_i)$ requires also the construction of the static control variate $h(x_i,y_i)$ defined by \eqref{eq:hxy}. This  depends on  the ratio $\widetilde{r}(x,y)$ given by \eqref{eq:tilde_ratio} and on the expectation $E_{q(y|x_i)}[h(x_i,y)]$. The calculation of the latter expectation is tractable since $\widetilde{r}(x,y)$ is the acceptance ratio of Metropolis-Hastings algorithm that targets the Gaussian target $\gengausspi(x) = N(x|\mu,\Sigma)$, where $\mu$ and $\Sigma$ are estimators of the mean and covariance matrix respectively of the target $\pi(x)$; see Assumption \ref{assumption1}.
 Finally, we compute $\hat{\theta}_n$ using \eqref{eq:theta_empirical} and evaluate the proposed estimator $\mu_{n,G}(F_{\hat{\theta}_n})$ from  \eqref{eq:estimator_final_theta}. Algorithm \ref{alg:VR} summarizes the steps of the  variance reduction procedure.

\begin{algorithm}[H]
\caption{Variance reduction for Metropolis-Hasting samplers}\label{alg:VR}
\textbf{Inputs}: The samples $x_i$, $i=0,\ldots,n-1$, simulated by using RWM or MALA with proposal distribution given by
\eqref{eq:qyx}; the proposed samples $y_i$ generated from the proposal during the MCMC; the M-H probabilities $\alpha(x_i,y_i)$ calculated during the MCMC; estimators $\mu$ and $\Sigma$ of the mean and covariance matrix respectively of the target.
\\
\textbf{Returns}: An estimate for the mean of the $j$th coordinate of the target.
\begin{algorithmic}[1]
\State Set $F(x) = x^{(j)}$.
\State Calculate $h(x_i,y_i)$ given by  \eqref{eq:hxy}.
\State Calculate $E_{q(y|x_i)}[h(x_i,y)]$ by utilising Propositions \ref{prop:PG} and \ref{prop:from_std_to_general_gauss}.
\State Calculate $\widehat{PG}(x_i)$ given by  \eqref{eq:PGhat} for each $i=1,\ldots,n$.
\State Calculate $\hat{\theta}_n$ given by \eqref{eq:theta_empirical}.
\State Return $\mu_{n,G}(F_{\hat{\theta}_n})$ given by \eqref{eq:estimator_final_theta}.
\end{algorithmic}
\end{algorithm}

%% file: Applications.tex
\section{Application on real and simulated data}
\label{sec:applications}

We present results from the application of the proposed methodology on real and simulated data examples. First we consider multivariate Gaussian targets for which we have shown that the function $G$ in \eqref{eq:G2dim} allows the explicit calculation of the expectation $PG$ defined by \eqref{eq:PG2}. Section \ref{sec:Gauss_sim} presents variance reduction factors in the case of $d$-variate standard Gaussian densities, simulated by employing the RWM and MALA, up to $d=100$ dimensions. In Sections \ref{sec:mix_gauss_sim}, \ref{sec:real_logistic} and \ref{sec:sv_sim} and  we examine the efficiency of our proposed methodology in targets that depart from the Gaussian distribution and the expectation $PG$ is not analytically available. 

To conduct all the experiments we set the parameters $b_0,b_1,b_2,c_0,c_1$ and $c_2$ of the function $G_0$ in \eqref{eq:G2dim} in the values given by Table \ref{tab:optimal} which were estimated by minimizing the loss function in \eqref{eq:loss} for $d=2$. 
In practice we observe that such values 
lead to good performance across all real data experiments, including those with $d>2$. 

To estimate the variance of $\mu_{n}(F)$ in each experiment we obtained $T=100$ different  estimates $\mu_{n}^{(i)}(F)$, $i=1,\ldots,T$, for $\mu_{n}(F)$ based on $T$ independent MCMC runs. Then, the variance of 
$\mu_{n}(F)$ has been estimated by 
$$
\frac{1}{T-1}\sum_{i=1}^{T}\{\mu_{n}^{(i)}(F)-\bar {\mu}_{n}(F)\}^2,
$$
where $\bar {\mu}_{n}(F)$ is the average of $\mu_{n}^{(i)}(F)$. We estimated similarly the variance of the proposed estimator $\mu_{n,G}(F)$.

\begin{table}[H]
\caption{Optimal values for the parameters of the function $G_0$ in \eqref{eq:G2dim}.}
\centering
\begin{tabular}{|ccccccc|}
 \hline
 & $b_0$ &$b_1$ &$b_2$ & $c_0$&$c_1$&$c_2$  \\
 \hline
RWM  &8.7078&	0.2916&	0.0001&	-3.5619& 0.1131&	3.9162\\
MALA  &7.6639& 0.0613&	0.0096&	-14.8086& 0.3431&	-0.0647
\\
\hline
\end{tabular}%
\label{tab:optimal}
\end{table}

\subsection{Simulated data: Gaussian targets}
\label{sec:Gauss_sim}

\begin{figure}[H]
    \centering
    \includegraphics[width=0.9\textwidth]{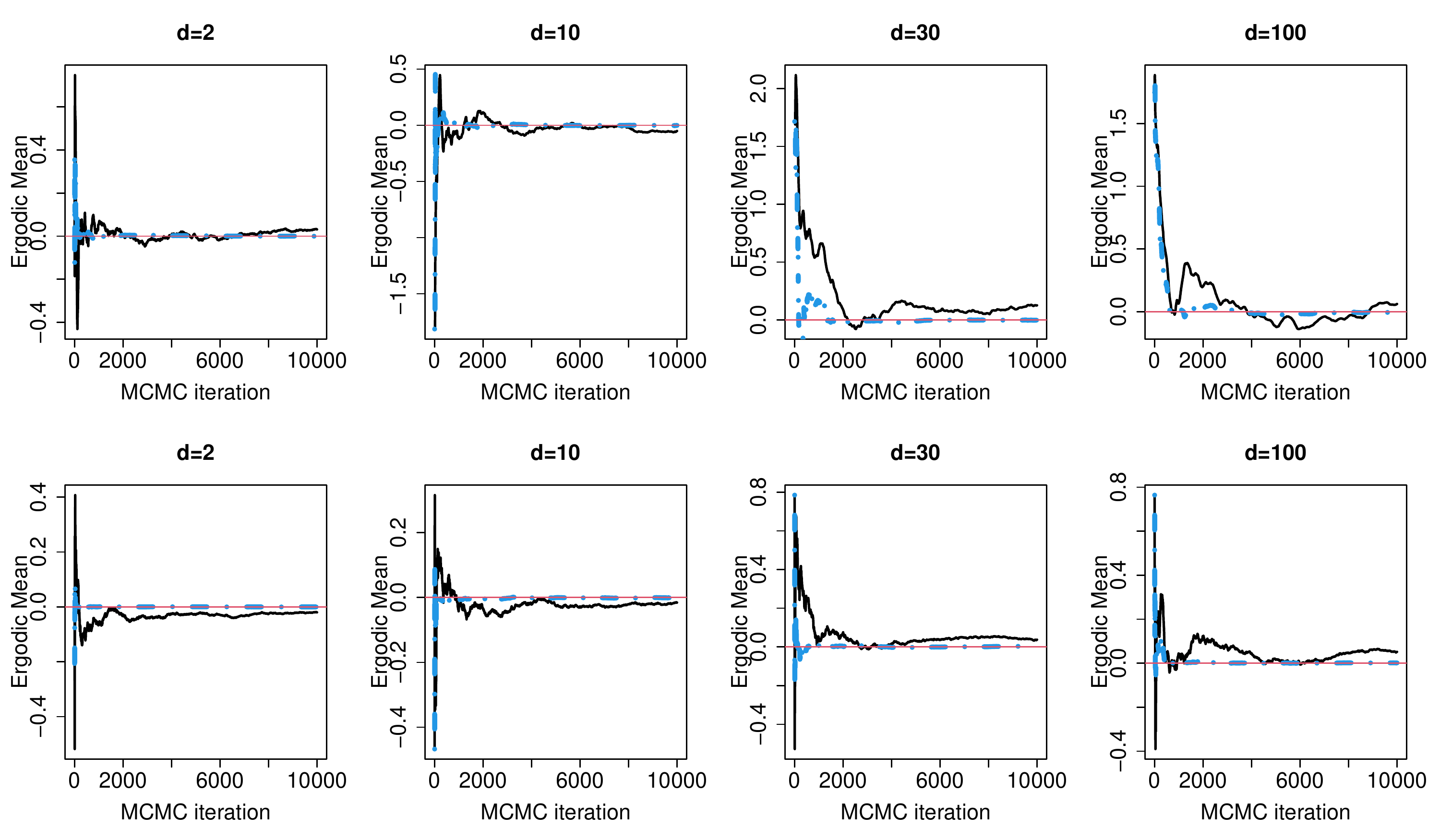}
    \vspace{-0.2cm}
     \caption{Sequence of the standard ergodic averages (black solid lines) and the proposed estimates (blue dashed lines). The red lines indicate the mean of the $d$-variate standard Gaussian target. The values are based on samples drawn by employing either the RWM (top row) or the MALA (bottom row) with $10,000$ iterations discarded as burn-in period.}
    \label{fig:erg_mean}
\end{figure}

\begin{figure}[H]
    \centering
    \includegraphics[width=0.9\textwidth]{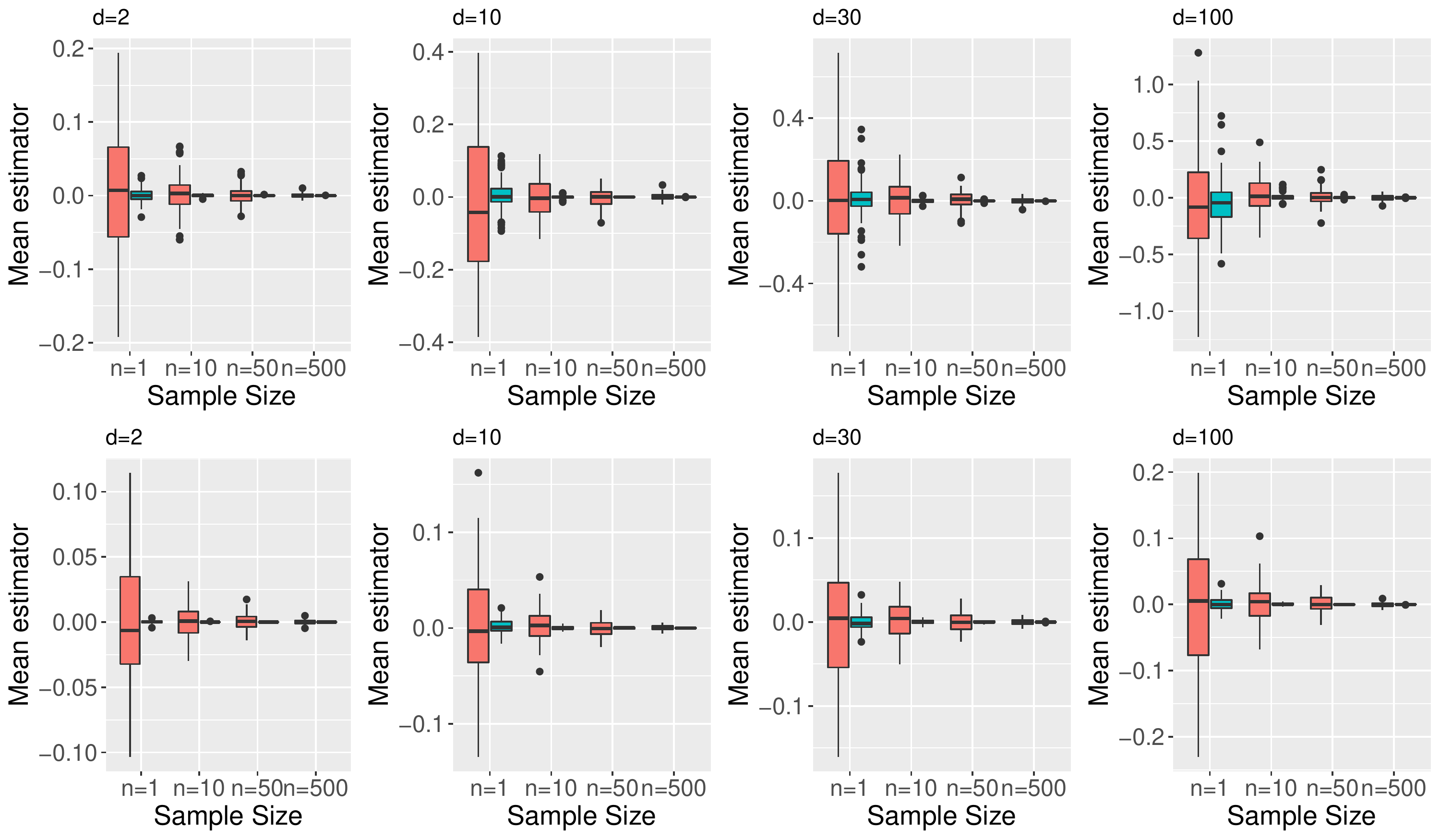}
    \vspace{-0.2cm}
    \caption{Each pair of boxplots is consisted of $100$ values for the estimators $\mu_n(F)$ (left boxplot) and $\mu_{n,G}(F_{\theta})$ (right boxplot) for the $d$-variate standard Gaussian target. The estimators have been calculated by using $n \times 10^3$ samples drawn by employing either the RWM (top row) or the MALA (bottom row) and discarded the first $10,000$ samples as burn-in period.}
    \label{fig:boxplots}
\end{figure}

The target distribution is a $d$-variate standard Gaussian distribution and we are interested in estimating the expected value of the first coordinate of the target by setting $F(x) = x^{(1)}$. 
Samples of size $n$ were drawn from target densities  by utilising the proposal distribution in \eqref{eq:qyx_std} with $c^2 = 2.38^2/d$ for the RWM case and by tuning $c^2$ during the burn-in period to achieve acceptance rate between $55\%$ and $60\%$ in the MALA case.
Table \ref{tab:std_Gaussian_var} presents factors by which the variance of $\mu_n(F)$ is greater than the variance of $\mu_{n,G}(F)$ in the case of the RWM and MALA.  Variance reduction is considerable even for $d=100$. Figure \ref{fig:erg_mean} shows typical realizations of the sequences of estimates obtained by the standard
estimators $\mu_n(F)$ and the proposed $\mu_{n,G}(F_{\theta})$ for different dimensions of the standard Gaussian target and Figure \ref{fig:boxplots} provides a visualization of the distribution of the estimators $\mu_n(F)$ and $\mu_{n,G}(F_{\theta})$.

\begin{table}[H]
\caption{Estimated factors by which the variance of $\mu_n(F)$ is larger than the variance of $\mu_{n,G}(F)$ for standard Gaussian $d$-variate target. We collect $n$ samples after the first $10,000$ iterations of the RWM and the MALA.}
\centering
\begin{tabular}{|lllllllll|}
\hline
&\multicolumn{4}{c}{RWM}&\multicolumn{4}{c|}{MALA}\\
\hline
 & d=2    & d=10   & d=30   & d=100& d=2    & d=10   & d=30   & d=100   \\
 \hline
n=1,000   & 93  & 26  & 10  & 5 & 1,345    & 64   & 57  & 97   \\
n= 10,000  & 278  & 173 & 112  & 27 & 3,572    & 81   & 88 & 316  \\
n= 50,000  & 541 & 445 & 177  &  94& 4,628   & 92   & 103   & 274  \\
n= 500,000 & 531 & 820 & 370 & 263& 4,997   & 83   & 157   & 286  \\
\hline
\end{tabular}
\label{tab:std_Gaussian_var}
\end{table}

\subsection{Simulated data: mixtures of Gaussian distributions}
\label{sec:mix_gauss_sim}

It is important to investigate how our proposed methodology performs when the target density departs from normality. We used as $\pi(x)$ a mixture of $d$-variate Gaussian distributions with density
\begin{equation}
    \label{eq:GaussMix}
   \pi(x) = \frac{1}{2}N(x|m,\Sigma) + \frac{1}{2}N(x|-m,\Sigma),
\end{equation}
where, following \cite{mijatovic2019asymptotic}, we set $m$ to be the $d$-dimensional vector $(h/2,0,\ldots,0)$ and $\Sigma$ is $d \times d$ covariance matrix randomly drawn from an inverse Wishart distribution by requiring its largest eigenvalue to be equal to $25$.

We drew samples from the target distribution by using the Metropolis-Hastings algorithm with proposal distribution $q(y|x) = N(y|x,c^2\Sigma)$ where by setting $c^2 = 2.38^2/d$ we achieve an acceptance ratio between $23\%$ and $33\%$.  When $h>6$ the MCMC algorithm struggles to converge.  Table \ref{tab:std_Gaussian_var_gaussmix} presents the factors by which the variance of $\mu_n(F)$ is greater than the variance of the modified estimator $\mu_{n,G}(F)$ for dimensions $d=10$ and $d=50$ and for different values of $h$.  It is very reassuring that even in the very non-Gaussian scenario $(h=6)$ our modified estimator achieved a slight variance reduction.

\begin{table}[H]
\caption{Estimated factors by which the variance of $\mu_n(F)$ is larger than the variance of $\mu_{n,G}(F)$ for a mixture of $d$-variate Gaussian distributions with density given by \eqref{eq:GaussMix} for different values of the mean $m$. We collect $n = 200,000$ samples after the first $10,000$ iterations of the RWM algorithm.}
\centering
\begin{tabular}{|llll|}
\hline
 &h=2& h=4& h=6\\
 \hline
d= 10& 20.73  & 2.39  &1.26   \\
d= 50&7.88  &1.35  &1.01   \\
\hline
\end{tabular}
\label{tab:std_Gaussian_var_gaussmix}
\end{table}

\subsection{Real data: Bayesian logistic regressions}
\label{sec:real_logistic}

We tested the variance reduction of our modified estimators on five datasets that have been commonly used in MCMC applications, see e.g. \cite{girolami2011riemann}, \cite{titsias2019gradient}. They are consisted of one $N$-dimensional binary response variable and an $N \times d$ matrix with covariates including a column of ones; see Table \ref{tab:logistic_data} for the names of the datasets and details on the specific samples sizes and dimensions. We consider a Bayesian logistic regression model by setting an improper prior for the regression coefficients $\gamma \in \mathbb{R}^d$ of the form $p(\gamma) \propto 1$. 

\begin{table}[H]
\caption{Summary of datasets for logistic regression}
\centering
\begin{tabular}{|lll|}
\hline
 Dataset &  d&  N  \\
 \hline
Ripley   & 3  & 250  \\
Pima Indian & 8 & 532 \\
Heart& 14 & 270\\
Australian  & 15  & 690  \\
German  & 25 & 1,000\\
\hline
\end{tabular}
\label{tab:logistic_data}
\end{table}

\subsubsection{Variance reduction for RWM }

We draw samples from the posterior distribution of $\gamma$ by employing the Metropolis-Hastings algorithm with proposal distribution 
\begin{equation*}
    \label{eq:logistic_proposal_rwm}
    q(\gamma'|\gamma) = N(\gamma'|\gamma,c^2\hat{\Sigma}),
\end{equation*}
where $c^2 = 2.38^2/d$ and $\hat{\Sigma}$ is the maximum likelihood estimator of the covariance of $\gamma$. Table \ref{tab:logistic_var_rwm} presents the range of factors by which the variance of $\mu_n(F)$ is greater than the variance of $\mu_{n,G}(F)$ for all parameters $\gamma$.  It is clear that our modified estimators achieve impressive variance reductions when compared with the standard RWM ergodic estimators.

\begin{table}[H]
\centering
\caption{Range of estimated factors by which the variance of $\mu_n(F)$ is larger than the variance of $\mu_{n,G}(F_{\theta})$ for the posterior distribution of logistic regression models applied on the datasets indicated by the first column. We collect $n$ samples after the first $10,000$ iterations of the RWM algorithm.}
\begin{tabular}{|llllll|}
\hline
  Dataset  & $n=1,000$& $n=10,000$ & $n=50,000$ & $n=200,000$     &  \\
\hline
Ripley    & 27.07-34.06 &  26.89-91.96&34.42-105.35 & 34.48-137.68 &  \\
Pima Indian   &14.62-25.91  &84.16-137.35 & 99.38-218.06 &  99.16-241.38&  \\
Heart   & 8.26-13.70 &16.63-40.81  &  23.53-64.07& 18.68-76.62 &  \\
Australian &6.14-15.27 &25.91-80.65  & 33.43-89.56 & 23.43-92.76&  \\
German &4.72-10.20 &19.61-54.63 &33.60-119.73 &25.61-148.54 &  \\
\hline
\end{tabular}
\label{tab:logistic_var_rwm}
\end{table}

\subsubsection{Variance reduction for MALA}

We draw samples from the posterior distribution of $\gamma$ by employing the Metropolis-Hastings algorithm with proposal distribution 
\begin{equation*}
    \label{eq:logistic_proposal_mala}
    q(\gamma'|\gamma) = N(\gamma'|\gamma + \tfrac{1}{2}c^2\hat{\Sigma}\nabla \log \pi(\gamma),c^2\hat{\Sigma} ), 
\end{equation*}
where $c^2$ is tuned during the burn-in period in order to achieve an acceptance ratio between $55\%$ and $60\%$, $\hat{\Sigma}$ is maximum likelihood estimator of the covariance of $\gamma$ and $\pi(\gamma)$ denotes the density of the posterior distribution of $\gamma$. Table \ref{tab:logistic_var_mala} presents the range of factors by which the variance of $\mu_n(F)$ is greater than the variance of $\mu_{n,G}(F)$ for all parameters $\gamma$. Again, there is considerable variance reduction for all modified estimators.

\begin{table}[H]
\centering
\caption{Estimated factors by which the variance of $\mu_n(F)$ is larger than the variance of $\mu_{n,G}(F_{\theta})$ for the posterior distribution of logistic regression models applied on the datasets indicated by the first column. We collect $n$ samples after the first $10,000$ iterations of the MALA.}
\begin{tabular}{|llllll|}
\hline
  Dataset  & $n=1,000$& $n=10,000$ & $n=50,000$ & $n=200,000$     &  \\
  \hline
Ripley    &10.89-15.99  &14.83-24.76  &  12.13 -26.06& 10.38 -20.14 &  \\
Pima Indian    & 23.50-51.64 & 34.95-52.42& 34.78-73.51&  36.64-72.75&  \\
Heart   &  10.04-17.31&   7.74-18.36&  10.11-18.07& 11.83-20.51  &  \\
Australian &9.32-22.78 &8.56-22.92  &   6.86-18.93&7.45-21.74 &  \\
German& 11.79-33.29& 11.39-42.46&9.80-52.24 & 8.21-40.72&  \\
\hline
\end{tabular}
\label{tab:logistic_var_mala}
\end{table}

\subsection{Simulated data: a stochastic volatility model}
\label{sec:sv_sim}

We use simulated data from a standard stochastic volatility model often employed in econometric applications to model the evolution of asset prices over time \citep{kim1998stochastic,kastner2014ancillarity}. By denoting with $r_t$, $t=1,\ldots,N$, the $t$th observation (usually log-return of an asset) the model assumes that $r_t =\exp\{h_t/2\}\epsilon_t$, where $\epsilon_t \sim N(0,1)$ and $h_t$ is an autoregressive AR(1) log-volatility, process: $h_t = m +\phi(h_{t-1}-m) + s\eta_t$, $\eta_t \sim N(0,1)$ and $h_0 \sim N(m,s^2/(1-\phi^2))$. To conduct Bayesian inference for the parameters $m \in \R$, $\phi \in (-1,1)$ and $s^2 \in (0,\infty)$ we specify commonly used prior distributions \citep{kastner2014ancillarity,alexopoulos2021bayesian}: $m \sim N(0,10)$, $(\phi+1)/2 \sim Beta(20,1/5)$ and $s^2 \sim Gam(1/2,1/2)$. The posterior of interest is 
\begin{align}
\label{eq:sv_target}
   \pi(m,\phi,s^2,h) &= p(m,\phi,s^2,h|r) \nonumber \\ & \propto p(m)p(s^2)p(\phi) N(h_0|m,s^2/(1-\phi^2))\prod_{t=1}^N N(r_t|0,e^{h_t})N(h_t|m+\phi(h_{t-1}-m),s^2), 
\end{align}
where $h=(h_0,\ldots,h_N)$ and $r=(r_1,\ldots,r_N)$.  

To assess the proposed variance reduction methods 
we simulated daily log-returns of a stock for $d$ days by using values for the parameters of the model that have been previously estimated in real data applications \citep{kim1998stochastic,alexopoulos2021bayesian} $\phi=0.98$, $\mu=-0.85$ and $s= 0.15$.  To draw samples from the $d$-dimensional, $d=N+3$, target posterior in \eqref{eq:sv_target} we first transform the parameters $\phi$ and $s^2$ to  real-valued parameters $\tilde{\phi}$ and $\tilde{s}^2$ by taking the logit and logarithm transformations and we assign Gaussian prior distributions by matching the first two moments of the Gaussian distributions with the corresponding moments of the beta and gamma distributions used as priors for the parameters of the original formulation. Then, we set $x=(m,\tilde{\phi},\tilde{s}^2,h)$ and we draw the desired samples using a Metropolis-Hastings algorithm with proposal distribution 
$$
q(y|x) = N(y|x + \tfrac{c^2}{2}\hat{\Sigma}\nabla\log\pi(x),c^2\hat{\Sigma} ),
$$
where $y= (m',\tilde{\phi}',\tilde{s}^{2'},h')$ are the proposed values, $c^2$ is tuned during the burn-in period in order to achieve an acceptance ratio between $55\%$ and $60\%$ and $\hat{\Sigma}$ is the maximum a posteriori estimate of the covariance matrix of $(m,\phi,s^2,h)$. Table \ref{tab:sv} presents the factors by which the variance of $\mu_n(F)$ is greater than the variance of the proposed estimator $\mu_{n,G}(F_{\theta})$.  We report variance reduction for all static parameters of the volatility process and the range of reductions achieved for the $N$-dimensional latent path $h$.   All estimators have achieved considerable variance reduction.

\begin{table}[H]
\caption{Estimated factors by which the variance of $\mu_n(F)$ is larger than the variance of $\mu_{n,G}(F)$ for the parameters of $d$-dimensional stochastic volatility model. We collect $n$ samples after the first $10,000$ of the MALA.}
\centering
\resizebox{\columnwidth}{!}{%
\begin{tabular}{|ccccc|cccc|cccc|}
\hline
 &\multicolumn{4}{c|}{$n=10,000$}& \multicolumn{4}{c|}{$n=50,000$}& \multicolumn{4}{c|}{$n=200,000$} \\ 
 \hline
 & $h$ &$m$ &$\phi$ & $s^2$&$h$ &$m$ &$\phi$ & $s^2$&$h$ &$m$ &$\phi$ & $s^2$ \\
 \hline
d=50 & 7.18-15.24 & 14.07& 17.44& 4.36& 7.49-15.92&13.02 &19.27&2.86&7.46-16.36 &16.97 &14.16 &2.43   \\
d=100  &1.06-7.66 & 7.99& 3.46& 1.26& 1.09-7.02&6.50 & 7.80&1.69 &1.52-9.54 &5.40 &4.44 &1.07  \\
\hline
\end{tabular}%
}
\label{tab:sv}
\end{table}

%% file: Discussion.tex
\section{Discussion}

Typical variance reduction strategies for MCMC algorithms study ways to produce new estimators which have smaller variance than the standard ergodic averages by performing a post-processing manipulation of the drawn samples. Here we studied a methodology that constructs such estimators but our development was based on the essential requirement of a negligible post-processing cost. In turn, this feature allows the effortless variance reduction for MCMC estimators that are used in a wide spectrum of Bayesian inference applications.

We investigated both the applicability of our strategy in high dimensions and the robustness to departures of normality in the target densities by using simulated and real data examples. Since we have never encountered a case in which variance increases, we feel that there is   strong evidence that our method is risk-free at least for posterior densities up to 100 dimensions. 

There are many directions for future work.  We limited ourselves to the simplest case of function $F(x) = x^{(j)}$ but higher moments and indicator functions seem interesting avenues to be investigated next.  Other Metropolis samplers such as the independent Metropolis or the Metropolis-within-Gibbs are also obvious candidates for studying.  Finally, an issue that was  discussed in some detail in \cite{dellaportas2009notes} but has not yet studied with the care it deserves is the important problem of reducing the estimation bias of the MCMC samplers which depends on the initial point of the chain $X_0 = x$ and vanishes asymptotically. As also noted by \cite{dellaportas2009notes}, control variables have probably an important role to play in this setting.

%% file: appendix_supplement.tex
\begin{center}
{\LARGE APPENDIX}
\end{center}

\section*{Proof of Proposition \ref{prop:PG}}
\begin{proof}
We need to calculate the integrals $a(x)$ and $a_g(x)$ in Eq. \eqref{eq:PGGaussian1} and \eqref{eq:PGGaussian2} for $G_0(x)$ given by \eqref{eq:G2dim}. We have that for $q(y|x)$ given by \eqref{eq:qyx_std}
\begin{equation*}
\label{eq:alpha_ratio_1_std_appdx}
\exp\left\{ -\tfrac{1}{2}(y^\top y - x^\top x)\right\} \frac{q(x|y)}{q(y|x)}  = \exp\big\{ -\tfrac{\tau^2}{2}(y^\top y-x^\top x)  \big\},
\end{equation*}
where $\tau^2=1$ in the case of RWM and $\tau^2=c^2/4$ in the case of MALA.

To compute $a(x)$ we set $z=(y-rx)/c$, where $r$ as in \eqref{eq:qyx_std}. Then, we have that  
\begin{equation}
\label{eq:alpha_ratio_3_std}
\exp\left\{ -\tfrac{\tau^2}{2}(y^\top y - x^\top x)\right\}  = \exp\big\{ -\tfrac{\tau^2c^2}{2}[ (z+\kappa)^\top (z+\kappa) -x^\top x/c^2   ]  \big\},
\end{equation}
where $\kappa = rx/c$. By setting $f= (z+\kappa)^\top (z+\kappa)$ we have that 
$f$ follows the non-central chi-squared distribution with $d$ degrees of freedom and non-central parameter $r^2x^\top x/c^2$. Eq. \eqref{eq:alpha_ratio_3_std} implies that $\alpha(x)$ in \eqref{eq:PGGaussian1} becomes
\begin{align}
\label{eq:min_computation}
a(x) &= \int \min\{1,\exp\{-\tfrac{\tau^2c^2}{2}[ f -x^\top x/c^2\}\}p(f)df \nonumber \\
& = \E_{f}\big[\min\{1, e^{-\tfrac{c^2\tau^2}{2}(f-x^\top x/c^2)}\}\big] \nonumber \\
& = \int_{-\infty}^{\tfrac{x^\top x}{c^2}} e^{-\tfrac{c^2\tau^2}{2}(f-x^\top x/c^2)}p(f)df + \int^{\infty}_{\tfrac{x^\top x}{c^2}}p(f)df
\end{align}
where $p(f)$ is the density of the random variable $f$ and writes
\begin{equation*}
\label{eq:pf_std}
p(f) = \sum_{j=0}^{\infty}\mathrm{Pois}\big(j| \kappa/2  \big)\mathrm{Gam}(f;d/2+j,2).
\end{equation*}
Notice that the second term in \eqref{eq:min_computation} can be calculated by using the cdf of the non-central chi squared distribution. For the first term after some algebra we have that 
\begin{equation}
\label{eq:pf_std2}
e^{-\tfrac{c^2\tau^2}{2}(f-x^\top x/c^2)}p(f) = \sum_{j=0}^{\infty}\frac{\mathrm{Pois}\big(j| \kappa/2  \big)}{(c^2\tau^2+1)^{d/2+j}}\mathrm{Gam}\big(f;d/2+j,\tfrac{2}{c^2\tau^2+1}\big).
\end{equation}

To compute $a_g(x)$ we first note that 
\begin{equation}
    \label{eq:compute_ag}
    a_g(x)  = \sum_{k=1}^K w_k a_{g_k}(x),
\end{equation}
where 
$$
a_{g_k}(x) = \int \min \left\{1, \exp\left\{ - \tfrac{1}{2}(y^\top y- x^\top x)\right\} 
\frac{q(x|y)}{q(y|x)} \right\}
g_k(y) q(y|x) dy,
$$
and $g_{k}(x) = \exp\{\beta_k^\top x - \gamma_k (x-\delta_k)^\top (x-\delta_k)\}$. 

Then, we calculate the $a_{g_k}(x)$ by noting that 
\begin{align*}
\label{eq:new_prop}
g_k(y)q(y|x) =  A_k(x)N\big(y|m_k(x),s_k^2\big),
\end{align*}
where 
$$
A_k(x) =(1+2c^2\gamma_k)^{-d/2}  \exp\bigg\{-\frac{r^2x^\top x}{2c^2}-\gamma_k\delta_k^\top \delta_k + \frac{m_k(x)^\top m_k(x)}{2c^2(1+2\gamma_kc^2)} \bigg\},
$$ 
$m_k(x) = \dfrac{rx + c^2(\beta_k+\gamma_k\delta_k)}{1+2c^2\gamma_k}$ and $s_k^2=c^2/(1+2c^2\gamma_k)$. By setting $z_{k,g} = (x-m_k(x))/s$ and $f_{k,g}= (z_{k,g}+\zeta_{k,g})^\top (z_{k,g}+\zeta_{k,g})$, where $\zeta_{k,g}= m_k(x)/s_k$, we work as in \eqref{eq:alpha_ratio_3_std} and have that
\begin{align*}
    a_{g_k}(x) &= A_k(x)\int \min\bigg\{1,\exp\{-\tfrac{\tau^2s_k^2}{2}[ f_{k,g} -x^\top x/s_k^2]\}\bigg\}p(f_{k,g})df_{k,g}\\
    & =A_k(x) \E_{f_{k,g}}\big[\min\{1, \exp\{-\tfrac{\tau^2s^2_k}{2}(f_{k,g}-x^\top x/s_k^2)\}\}\big],
\end{align*}
where the random variable $f_{k,g}$ follows the chi-squared distribution with $d$ degrees of freedom and non-central parameters $m_k(x)^\top m_k(x)/s_k^2$ and the expectation is calculated by utilizing the cdf of $f_{k,g}$ as in Eq. \eqref{eq:min_computation}-\eqref{eq:pf_std2}. Finally, from Eq. \eqref{eq:compute_ag} we have that 
\begin{equation}
\label{eq:agx_final}
a_g(x) = \sum_{k=1}^K A_k(x) \E_{f_{k,g}}\big[\min\{1, \exp\{-\tfrac{\tau^2s^2_k}{2}(f_{k,g}-x^\top x/s_k^2)\}\}\big].
\end{equation}

\end{proof}

\section*{Proof of Proposition \ref{prop:properties_of_stdgausssol}}

Let $q(y|x)$ be the proposal distribution defined by \eqref{eq:qyx_std}. We have that 
$$
\frac{\widetilde{\pi}_0(y)q(x|y)}{\widetilde{\pi}_0(x)q(y|x)} = \exp\{-\frac{\tau^2}{2}(y^\top y-x^\top x)\},
$$
where $\tau^2=1$ in the case of RWM and $\tau^2=c^2/4$ in the case of MALA. We assume that $F(x) = x^{(j)}$ and we show that i) $\hat{F}^0_{\widetilde{\pi}}(-x^{(j)},x^{(j')}) = -\hat{F}^0_{\widetilde{\pi}}(x)$ and that ii) $\hat{F}^0_{\widetilde{\pi}}(x^{(j)},x^{(j')}) = \hat{F}^0_{\widetilde{\pi}}(x^{(j)},\Pi x^{(j')})$, where $x^{(j')}$ denotes the vector $x \in \R^d$ without its $j$th coordinate and $\Pi$ is a permutation matrix.

Since $\hat{F}^0_{\widetilde{\pi}}$ satisfies the Poisson equation we have that
\begin{equation}
\label{eq:std_gauss_pois_appendix}
 \int \min\big\{1,\exp\{-\frac{\tau^2}{2}(y^\top y-x^\top x)\}\big\}[\hat{F}^0_{\widetilde{\pi}}(x)-\hat{F}^0_{\widetilde{\pi}}(y)]q(y|x)dy = x^{(j)},
\end{equation}
which implies that 
\begin{equation}\label{eq:std_gauss_pois_appendix_2}
-\hat{F}^0_{\widetilde{\pi}}(x)\int  \alpha(x,y)q(y|x)dy   - \int  \alpha(x,y)[-\hat{F}^0_{\widetilde{\pi}}(y)]q(y|x)dy = -x^{(j)},
\end{equation}
where $\alpha(x,y) = \min\big\{1,\exp\{-\frac{\tau^2}{2}(y^\top y-x^\top x)\}\big\}$. Let also $z$ and $\tilde{z}$ be $d$-dimensional vectors such that $z^{(j)}=-x^{(j)}$ and $z^{(-j)}=x^{(-j)}$ and $\tilde{z}^{(j)}=-y^{(j)}$, $\tilde{z}^{(j')}=y^{(j')}$. 
Then, by noting that the Jacobian of the transformations is equal to one, \eqref{eq:std_gauss_pois_appendix_2} becomes 
\begin{equation}\label{eq:std_gauss_pois_appendix_3}
-\hat{F}^0_{\widetilde{\pi}}(-z^{(j)},z^{(j')}) \int  \alpha(z,\tilde{z})q(\tilde{z}|z)d\tilde{z}   - \int  \alpha(z,\tilde{z})[-\hat{F}^0_{\widetilde{\pi}}(-z^{(j)},z^{(j')})]q(\tilde{z}|z)d\tilde{z} = z^{(j)},
\end{equation}
where $\alpha(z,\tilde{z}) = \min\big(1,e^{-\frac{c^2+r^2-1}{2c^2}(\tilde{z}^\top\tilde{z}-z^\top z)}\big)$ and $q(\tilde{z}|z)=N(\tilde{z}|rz,c^2I)$. Equation \eqref{eq:std_gauss_pois_appendix_3} implies that $-\hat{F}^0_{\widetilde{\pi}}(-z^{(j)},z^{(j')})$ is solution of the Poisson equation and from the uniqueness of the solution we have i). 

To prove ii) we denote by $z$ the $d$-dimensional vector such that $z^{(j)} = x^{(j)}$ and $z^{(j')} = \Pi x^{(j')}$ and we apply the following transformation on \eqref{eq:std_gauss_pois_appendix}; we set $\tilde{z}$ to be $d$-dimensional vector such that $\tilde{z}^{(j)} = y^{(j)}$ and $\tilde{z}^{(j')} = \Pi y^{(j')}$. Then, we have that 
\begin{equation}\label{eq:std_gauss_pois_appendix_4}
\hat{F}^0_{\widetilde{\pi}}(z^{(j)},\Pi^{-1}z^{(j')}) \int  \alpha(z,\tilde{z})q(\tilde{z}|z)d\tilde{z}   - \int  \alpha(z,\tilde{z})[\hat{F}^0_{\widetilde{\pi}}(z^{(j)},\Pi^{-1}z^{(j')})]q(\tilde{z}|z)d\tilde{z}= z^{(j)},
\end{equation}
where $\alpha(z,\tilde{z})$ and $q(\tilde{z}|z)$ as in \eqref{eq:std_gauss_pois_appendix_3} since they are invariant to arbitrary permutations of $z$ and/or $\tilde{z}$ and $\Pi^{-1}$ is permutation matrix such that $\Pi^{-1}\Pi z^{(j')} = z^{(j')}$. From \eqref{eq:std_gauss_pois_appendix_4} we have that $\hat{F}^0_{\widetilde{\pi}}(z^{(j)},\Pi^{-1}z^{(j')})$ is solution of the Poisson equation and then ii) holds again due to the uniqueness of the solution of the Poisson equation.

\section*{Proof of Proposition \ref{prop:from_std_to_general_gauss}}

\begin{proof}
Let $\widetilde{\pi}(z) =N(z|0,I)$ be the target of a Metropolis-Hastings algorithm with proposal $q(\tilde{z}|z) =N(\tilde{z}|rz,c^2I)$, it easy to see that $r=1$ corresponds to the RWM algorithm and  $r=1-c^2/2$ to MALA. 
Let also $\hat{F}^0_{\tilde{\pi}}$ the solution of the associated Poisson equation and $\widetilde{\pi}(x) =N(x|\mu,\Sigma)$ be a bivariate Gaussian density with mean $\mu$ and covariance matrix $\Sigma$. We assume without loss of generality that $F(z) = z^{(1)}$ in \eqref{eq: PE} which becomes
\begin{equation}
\label{eq:pois_eq_std}
\Bigg( \int \alpha_z(z,\tilde{z})q(\tilde{z}|z)d\tilde{z} \Bigg)\hat{F}^0_{\tilde{\pi}}(z)- \int \alpha_z(z,\tilde{z})q(\tilde{z}|z)   \hat{F}^0_{\tilde{\pi}}(\tilde{z})d\tilde{z} = z^{(1)},
\end{equation}
where 
\begin{equation}
\label{eq:min_std}
\alpha_z(z,\tilde{z}) = \min \bigg\{1, e^{-\tfrac{c^2+r^2-1}{2c^2}(\tilde{z}^\top\tilde{z}-z^\top z)}\bigg\}
\end{equation}
Let $x= \mu + Lz$ and $y= \mu + L\tilde{z}$, where $L$ such that $\Sigma=LL^\top$. From the properties of the Gaussian distribution we  have that 
\begin{align*}
    q(y|x) &= N( y| r (x-\mu) +\mu, c^2\Sigma  )
\end{align*}
Moreover, equation \eqref{eq:min_std} becomes
\begin{equation*}
\label{eq:new_acc}
\alpha_z\big(L^{-1}(x-\mu),L^{-1}(y-\mu)\big) = \min \big\{1, e^{-\tfrac{c^2+r^2-1}{2c^2}\big[(y-\mu)^\top\Sigma^{-1}(y-\mu)-(x-\mu)^\top\Sigma^{-1}(x-\mu)\big]}\big\} =  \widetilde{\alpha}(x,y)
\end{equation*}
Then, equation \eqref{eq:pois_eq_std} becomes 
\begin{equation}
\label{eq:pois_eq_std_trans}
\Bigg( \int \widetilde{\alpha}(x,y)q(y|x)dy \Bigg)\hat{F}^0_{\tilde{\pi}}\big(L^{-1}(x-\mu)\big)- \int \widetilde{\alpha}(x,y)q(y|x)   \hat{F}^0_{\tilde{\pi}}\big(L^{-1}(x-\mu)\big)dy = L^{-1}_{11}(x^{(1)}-\mu^{(1)}),
\end{equation}
where $L_{11}$ is the first diagonal element of $L$.
Since $L_{11}=1/L^{-1}_{11}$ Equation \eqref{eq:pois_eq_std_trans} implies that the function 
$$
\hat{F}_{\tilde{\pi}}(x) =L_{11}\hat{F}^0_{\tilde{\pi}}\big(L^{-1}(x-\mu)\big)
$$
is the solution of the Poisson equation associated to the Metropolis-Hastings algorithm with target $\tilde{\pi}(x)$ and proposal $q(y|x)$.

\end{proof}

\section*{Calculations for Remark \ref{rem:MALA_static}}

As noted in Section \ref{sec:static_control} the calculation of $E_q[h(x,y)]$ requires the to compute the following integral
\begin{align}
\label{eq:comp_eqh}
     \int \min\left\{1,\exp\{- \frac{1}{2}(\tilde{y}^\top \tilde{y} - \tilde{x}^\top \tilde{x})  \}\frac{\widetilde{q}(\tilde{x}|\tilde{y})}{\widetilde{q}(\tilde{y}|\tilde{x})} \right\} \big[\stdgaussG(\tilde{y})- \stdgaussG(\tilde{x}) \big]q(\tilde{y}|\tilde{x})  d \tilde{y}, 
\end{align}
where $\tilde{x}$ and $\tilde{y}$ as defined in Section \ref{sec:static_control}. In the case of the RWM algorithm the calculation of the integral above is conducted by using the results in Proposition \ref{prop:PG} since the the densities $\widetilde{q}(\tilde{y}|\tilde{x})$ and $q(\tilde{y}|\tilde{x})$ coincide and, thus, \eqref{eq:comp_eqh} is consisted of the integrals in \eqref{eq:PGGaussian1} and \eqref{eq:PGGaussian2}. 

In the case of the MALA $\widetilde{q}(\tilde{y}|\tilde{x})$, which is given by \eqref{eq:qyx_std}, has mean $r\tilde{x}$ whereas the mean of the distribution with density $q(\tilde{y}|\tilde{x})$ in \eqref{eq:qyx_reparam} is $k(\tilde{x}) = \tilde{x}+(c^2/2)L^\top\nabla\log \pi(x) $. However, the calculation of 
$$
a^h(x) = \int \min\left\{1,\exp\{- \frac{1}{2}(\tilde{y}^\top \tilde{y} - \tilde{x}^\top \tilde{x})  \}\frac{\widetilde{q}(\tilde{x}|\tilde{y})}{\widetilde{q}(\tilde{y}|\tilde{x})}dy \right\}q(\tilde{y}|\tilde{x}) 
$$
is conducted similarly to the calculation of $a(x)$ in the Proof of Proposition $2$ and, more precisely, we have that $a^h(x)$ is calculated from equation \eqref{eq:min_computation} where $f$ follows the chi-squared distribution with $d$ degrees of freedom and non-central parameter $k(\tilde{x})^\top k(\tilde{x})/c^2$.  To compute the integral  
$$
a^h_g(x) = \int \min\left\{1,\exp\{- \frac{1}{2}(\tilde{y}^\top \tilde{y} - \tilde{x}^\top \tilde{x})  \}\frac{\widetilde{q}(\tilde{x}|\tilde{y})}{\widetilde{q}(\tilde{y}|\tilde{x})}\stdgaussG(\tilde{y})dy \right\} 
$$
we work again as in the proof of Proposition $2$ for the calculation of $a_g(x)$ and we find that $a_g^h(x)$ is given by equation \eqref{eq:agx_final} for 
$$
A_k(x) = (1+2c^2\gamma_k)^{-d/2}  \exp\bigg\{-\frac{k(\tilde{x})^\top k(\tilde{x})}{2c^2}-\gamma_k\delta_k^\top \delta_k + \frac{m_k(x)^\top m_k(x)}{2c^2(1+2\gamma_kc^2)} \bigg\}
$$
and
$$
m_k(x) = \dfrac{k(x) + c^2(\beta_k+\gamma_k\delta_k)}{1+2c^2\gamma_k}.
$$